\documentstyle[preprint,eqsecnum,aps]{revtex}
\begin{document}
\draft
\title{Localized electronic states and photoemission\\
superconducting condensate in Bi$_{2}$Sr$_{2}$Ca$_{1}$Cu$_{2}$O$_{8+x}$}
\author{Jian Ma,$^{a}$ C.Quitmann,$^{a}$ R.J.Kelley,$^{a}$\\
G. Margaritondo,$^{b}$ and M.Onellion$^{a}$}
\address{$^{a}$Department of Physics and Applied Superconductivity Center,
University of Wisconsin-Madison, Madison, WI 53706\\
$^{b}$Institut de Physique Appliqu\'{e}e, Ecole Polytechnique F\'{e}d\'{e}rale,
CH-1015 Lausanne, Switzerland}
\date{\today}
\maketitle

\begin{abstract}
We present the first detailed angle-resolved photoemission evidence that there
are two types of carriers that contribute to the photoemission superconducting
condensate in $Bi_{2}Sr_{2}Ca_{1}Cu_{2}O_{8+x}$. Our data indicate that
both itinerant and somewhat localized normal state carriers can contribute
to the formation of Cooper pairs.
\end{abstract}

%\section{Introduction}
One of the long-standing controversies in the cuprate superconductors is
the nature of the normal state. In addition to the by now classic linear
resistivity up to high temperatures,\cite{Batlogg} the changes with
stoichiometry from underdoped to overdoped remain difficult to fit into a
single picture.
Among the most difficult matters has been whether there are one or two types
of carriers.\cite{MaLee,Mills,Heeger,Samokhvalov,Rentschler,Bianconi,%
Egami,Basov}
X-ray absorption\cite{Bianconi}, neutron scattering\cite{Egami} and ab-plane
optical conductivity\cite{Basov} studies indicate that there are somewhat
localized carriers, with a wavefunction diameter of about 1 nm. However,
there has not been direct spectroscopic evidence that
such carriers exist, nor that such carriers contribute to the superconducting
condensate. In this report, we provide such data and discuss the implications.
To avoid misleading the reader, we emphasize that the samples on which we
report are exceptional. Most of our samples ($>$95\%) that exhibit a
superconducting condensate also exhibit an itinerant band in the normal state.
However, there are a small number of samples (about 3\%) that exhibit a
superconducting condensate in the absence of a normal state itinerant band.
This report concentrates on these exceptional samples.

%\section{Experimental}
Synchrotron-radiation angle-resolved photoemission experiments were performed
at the Wisconsin Synchrotron Radiation Center. The details of the experimental
procedure are provided elsewhere.\cite{Kelley1,Ma} We used a 50 mm
hemispherical
electron energy analyzer, mounted on a two-axis goniometer. The samples
were transferred through a load-lock chamber and cleaved in {\it situ}, at 35K,
in a vacuum of $8 \times 10^{-11}$
torr. Photoemission spectra were taken below the superconducting transition
temperature, $T_{c}$ = 83K. We then raised the sample temperature to 95K,
and took normal state photoemission spectra.

%\section{Results and Discussion}
Figures\ \ref{GammaY}-\ref{GammaM} illustrate superconducting state and
normal state spectra taken
along the three major symmetry directions, including the Cu-O-Cu bond axis in
real space, and the Bi-O-Bi a-axis and b-axis.  The superconducting state
and normal state spectra in a given symmetry direction were taken on the same
sample; only the temperature is different.  Figures.\ \ref{GammaY}-\ref{GammaM}
provide what is to our
knowledge {\it the first spectroscopic evidence that a superconducting
condensate can be observed in the absence of a normal state itinerant band}.

Several noteworthy points emerge from Figs.\ \ref{GammaY}-\ref{GammaM}.
The size of the gap, as noted in the figure captions, is not the same in
the three high-symmetry directions. We define the energy position of the
midpoint of the leading edge of the superconducting state spectrum as the
gap. Our estimates of the gap size are:
$\Delta_{\Gamma-\bar{M}}$ = 20 meV, $\Delta_{\Gamma-Y}$ = 14 meV, and
$\Delta_{\Gamma-X}$ = 10 meV. Along the $\Gamma-\bar{M}$ and $\Gamma-X$
directions, the data of Figs.\ \ref{GammaX}-\ref{GammaM} yield estimates
of the gap size similar to the
estimates obtained from samples that exhibit a normal state itinerant band.
However, the two types of samples yield very different estimates of the gap
size along $\Gamma-Y$: $\Delta_{\Gamma-Y}$ = 14 meV (Fig.\ \ref{GammaY})
compared to 0-2 meV.\cite{Kelley2}

We compared the angular extent of the gap and condensate observed for
the data of Figs.\ \ref{GammaY}-\ref{GammaM} to samples that
exhibit a normal state itinerant band. Note that the gap in
Figs.\ \ref{GammaY}-\ref{GammaM} is observed for almost the exact same
locations in the Brillouin zone as for samples that exhibit a
normal state itinerant band. This result indicates that the superconducting
gap is forming near the Fermi surface, as it should, independent of the nature
of the carriers.

We observe no normal state itinerant band for the samples illustrated in
Figs.\ \ref{GammaY}-\ref{GammaM}. Since the normal state itinerant band would
have at most a factor of two reduction in spectral area (see above), the
absence of such a band is conclusively established.

There are at least two interpretations of our data. As Ma and Lee have
\cite{MaLee} noted, the results could be explained if the scattering in
the normal state was sufficiently strong, and the normal state scattering
was largely eliminated in the superconducting state. There are reports that
the normal state scattering rate of the cuprates is higher than elemental
metals. In addition, there are reports that below $T_{c}$, the
scattering rate of the quasiparticles remaining outside the condensate
drops dramatically.\cite{Bonn}  However, such an
interpretation does not fully account for the data in
Figs.\ \ref{GammaY}-\ref{GammaM}. The scattering
rate in question must be an inelastic scattering channel to be suppressed by
the opening of a superconducting gap. As the data in earlier
reports\cite{Olson1} make clear, however, the normal state quasiparticle
spectral feature has an energy width of 150 meV, much
larger than the superconducting gap. Thus, the opening of a superconducting
gap of about 25 meV will not suppress an inelastic scattering channel that must
lead to the removal of a normal state quasiparticle spectral feature having
an energy width of about 150 meV.

Instead, our data can be consistently interpreted in a simpler
way, {\it viz}, that
the samples have somewhat localized
carriers in the normal state that contribute to the photoemission
superconducting condensate spectral
feature. This interpretation is supported by other evidence as well. We have
recently reported\cite{Almeras} that cobalt-doping of Bi-2212 leads to
the three classic
characteristics of Anderson localization, including the removal of the normal
state quasiparticle spectral features due to localization. Such data indicates
that somewhat localized carriers still contribute to the superconducting state,
since a superconducting transition is observed in resistivity measurements,
for both cobalt-doped samples and the present samples.

Our data do not allow us to determine the nature of the somewhat localized
carriers. There have been several proposals in the literature on such
carriers. The speculations include a type of Peierls
distortion in two dimensions.\cite{Liu} In particular, note that the difference
between the superconducting and normal state spectra is not limited to
the spectral area of the photoemission superconducting condensate. Instead,
the data indicate that spectral area at higher binding energy than the
condensate appears in the superconducting state and is lost in the normal
state. Such a result follows from a picture in which the distortions within
the $CuO_{2}$ unit are randomly arranged above $T_{c}$ but become ordered
below $T_{c}$.
The random arrangement above $T_{c}$ means that the electron wavevector,
{\bf k}, is
not a good quantum number. Consequently, we would
not observe a distinct itinerant
quasiparticle band state. However, such models\cite{Liu} suggest that below
$T_{c}$ the electron wavevector becomes a good quantum number and the
quasiparticle band is therefore observed, as is the superconducting condensate.
Our data, while consistent with such an interpretation, do not conclusively
establish the model.

%\section{Conclusion}
In summary, we have observed a superconducting condensate, and spectral
features at higher binding energy, in the superconducting state, as one would
expect from the literature\cite{Ma,Kelley2,Olson2,Hwu,Shen} for typical
samples. Above $T_{c}$, however, we do
not observe any itinerant quasiparticle band. The data establish that both
itinerant and more localized carriers contribute to the condensate. We do not
yet have conclusive evidence on the nature of the more localized carriers.

\vspace{.8in}
%\noindent ACKNOWLEDGEMENT
We benefitted from conversations with Andrey Chubukov, Robert Joynt and
Peter Dowben. The staff of the Wisconsin Synchrotron Radiation Center (SRC),
particularly Robert Pedley, were most helpful. Financial support was
provided by the U.S. NSF, both directly(DMR-9214701) and through support of
the SRC, by Ecole Polytechnique F\'{e}d\'{e}rale Lausanne and the Fonds
National Suisse de la Recherche Scientifique, and by Deutsche
Forschungsgemeinschaft.

\eject

\begin{figure}
\caption{(a). Angle-resolved photoemission (ARUPS) spectra of the
superconducting state (35K), and the normal state (95K) along $\Gamma-Y$
direction. The gap opening at $\theta = 13^\circ$ is
$\Delta_{\Gamma-Y}$ = 14 meV. (b). ARUPS normal state spectra (95K) along
the same direction in the wider binding energy range. Note that no distinct
itinerant band feature is visible.}
\label{GammaY}
\end{figure}

\begin{figure}
\caption{(a). Angle-resolved photoemission (ARUPS) spectra of the
superconducting state (35K), and the normal state (95K) along $\Gamma-X$
direction. The gap opening at $\phi = 13^\circ$ is
$\Delta_{\Gamma-X}$ = 10 meV. (b). ARUPS normal state spectra (95K) along
the same direction in the wider binding energy range. Note that no distinct
itinerant band feature is visible.}
\label{GammaX}
\end{figure}

\begin{figure}
\caption{(a). Angle-resolved photoemission spectra of the superconducting
state (35K), and (b) the normal state (95K) along $\Gamma-\bar{M}$ direction.
The gap opening at $\theta/\phi = 18^\circ/18^\circ$ is
$\Delta_{\Gamma-\bar{M}}$ = 20 meV. Again, no normal state itinerant band
is visible.}
\label{GammaM}
\end{figure}


\begin{references}
\bibitem{Batlogg}For a review, see B. Batlogg, Physica B {\bf 169}, 7 (1991).

\bibitem{MaLee}M. Ma and P.A. Lee, Phys. Rev. B {\bf 32}, 5658 (1985).

\bibitem{Mills}A.J. Mills and Boris I. Shraiman, Phys. Rev. B
{\bf 46}, 14843 (1992).

\bibitem{Heeger}Alan J. Heeger and Gang Yu, Phys. Rev. B {\bf 48}, 6492 (1993).

\bibitem{Samokhvalov}A.A. Samokhvalov, N.A. Viglin, B.A. Gizhevskii,
N.N. Loshkareva, V.V. Osipov, N.I. Solin and Yu.P. Sukhorukov,
JETP {\bf 76} 463 (1993).

\bibitem{Rentschler}T. Rentschler, S. Kemmler-Sack, P. Kessler and H. Lichte,
Physica C {\bf 219}, 167 (1994).

\bibitem{Bianconi}A. Bianconi, S. Della Longa, M. Missori, C. Li,
M. Pompa, A. Soldatov, S. Turtu and S. Pagliuca, in the Proceedings of
the Beijing International Conference: High Temperature
Superconductivity (BHTSC 92), edited by Z.Z. Gan, S.S. Xie, Z.X. Zhao
(World Scientific, Singapore, 1992), p. 147-55.

\bibitem{Egami}S.J.L. Billinge and T. Egami, in Lattice Effects in
High $T_{c}$ superconductors, edited by Y. Bar-Yam, T. Egami,
J. Mustre-de-Leon, A.R. Bishop (World Scientific,Singapore, 1992),
p. 93-104; see also T. Egami, Intl. Conf. on Nano-Engineering and High-
Temperature Superconductors, I. Bozovic and D. Pavuna, eds., World Scientific
(1994).

\bibitem{Basov}D.N. Basov, A.V. Puchkov, R.A. Hughes, T. Strach, J. Preston,
T. Timusk, D.A. Bonn, R. Liang and W.N. Hardy, Phys. Rev. B {\bf 49},
12165 (1994).

\bibitem{Kelley1}R.J. Kelley, Jian Ma, G. Margaritondo and M. Onellion,
Phys. Rev. Lett. {\bf 71}, 4051 (1993).

\bibitem{Ma}Jian Ma, C. Quitmann, R.J. Kelley, G. Margaritondo, and
M. Onellion, submitted.

\bibitem{Kelley2}R.J. Kelley, Jian Ma, R. Joynt, H. Berger, G. Margaritondo,
and M. Onellion, submitted.

\bibitem{Bonn}D.A. Bonn, P. Dosanjh, R. Liang and W.N. Hardy, Phys. Rev.
Lett. {\bf 68}, 2390 (1992).

\bibitem{Olson1}C.G. Olson, R. Liu, D.W. Lynch, R.S. List, A. J. Arko,
B.W. Veal, Y.C. Chang, P.Z. Jiang, A.P. Paulikas, Phys. Rev. B {\bf 42},
381 (1990).

\bibitem{Almeras}P. Alm\'{e}ras, H. Berger, G. Margaritondo, Jian Ma,
C. Quitmann, R.J. Kelley, M. Onellion, submitted.

\bibitem{Liu}J.N. Liu, X. Sun, R.T. Fu and K.Nasu, Phys. Rev. B {\bf 46},
1710 (1992).

\bibitem{Olson2}C.G. Olson, R. Liu, A.-B. Yang, D.W. Lynch, A.J. Arko,
R.S. List, B.W. Veal, Y.C. Chang, P.Z. Jiang and A.P. Paulikas,
Science {\bf 245}, 731 (1989).

\bibitem{Hwu}Y. Hwu, L. Lozzi, M. Marsi, S. La Rosa, M. Winokur,
P. Davis, M. Onellion, H. Berger, F. Gozzo, F. L\'{e}vy
and G. Margaritondo, Phys. Rev. Lett., {\bf 67}, 2573 (1991).

\bibitem{Shen}Z.-X. Shen, D.S. Dessau, B.O. Wells, D.M. King, W.E. Spicer,
A.J. Arko, D. Marshall, L.W. Lombardo, A. Kapitulnik, P. Dickinson,
S. Doniach, J. DiCarlo, A.G. Loeser and C.H. Park, Phys. Rev. Lett.,
{\bf 70}, 1553 (1993).
\end{references}
\end{document}